\documentclass[aps,prd,secnumarabic,nobibnotes,twocolumn]{revtex4-1}
\usepackage{amsfonts}
\usepackage{mathrsfs}
\usepackage{amsmath}
\usepackage{color}
\usepackage{natbib}
\usepackage{graphicx}
\usepackage{bm}
\usepackage{amssymb}
\usepackage{xspace}
\usepackage{epstopdf}
\usepackage{dcolumn}
\usepackage{multirow}
\usepackage[colorlinks=true, letterpaper=true, pdfstartview=FitV, linkcolor=blue, citecolor=blue, urlcolor=blue]{hyperref}
\usepackage{wrapfig}

\makeatletter

\newcommand{\Rmnum}[1]{\expandafter\@slowromancap\romannumeral #1@}
\makeatother

\begin{document}
\title{Ferromagnetic hybrid nodal loop and switchable type-I and type-II Weyl fermions in two-dimension}

\author{Tingli He}
\address{School of Materials Science and Engineering, Hebei University of Technology, Tianjin 300130, China.}

\author{Xiaoming Zhang}\email{zhangxiaoming87@hebut.edu.cn}
\address{School of Materials Science and Engineering, Hebei University of Technology, Tianjin 300130, China.}

\author{Zhi-Ming Yu}
\address{School of Physics, Beijing Institute of Technology, Beijing 100811, China}

\author{Ying Liu}
\address{School of Materials Science and Engineering, Hebei University of Technology, Tianjin 300130, China.}

\author{Xuefang Dai}
\address{School of Materials Science and Engineering, Hebei University of Technology, Tianjin 300130, China.}

\author{Guodong Liu}\email{gdliu1978@126.com}
\address{School of Materials Science and Engineering, Hebei University of Technology, Tianjin 300130, China.}

\author{Yugui Yao}
\address{School of Physics, Beijing Institute of Technology, Beijing 100811, China}

\begin{abstract}
As a novel type of fermionic state, hybrid nodal loop with the coexistence of both type-I and type-II band crossings has attracted intense research interest. However, it remains a challenge to realize hybrid nodal loop in both two-dimensional (2D) materials and in ferromagnetic (FM) materials. Here, we propose the first FM hybrid nodal loop in 2D CrN monolayer. We show that the material has a high Curie temperature ($>600$ K) FM ground state, with the out-of-plane [001] magnetization. It shows a half-metallic band structure with two bands in the spin-up channel crossing each other near the Fermi level. These bands produce both type-I and type-II band crossings, which form a fully spin-polarized hybrid nodal loop. We find the nodal loop is protected by the mirror symmetry and robust against spin-orbit coupling (SOC). An effective Hamiltonian characterizing  the hybrid nodal loop is established. We further find the configuration of nodal loop can be shifted under external perturbations such as strain. Most remarkably, we demonstrate that both type-I and type-II Weyl nodes can be realized from such FM hybrid nodal loop by simply shifting the magnetization from out-of-plane to in-plane. Our work provides an excellent candidate to realize FM hybrid nodal loop and Weyl fermions in 2D material, and is also promising for related topological applications with their intriguing properties.
\end{abstract}
\maketitle

\section{Introduction}
Two-dimensional (2D) materials, best exemplified by graphene, have been attracting significant interest currently~\cite{1,2,3}. They not only show novel characteristics that distinct from the 3D bulk, but also can provide the opportunity to easily control their characteristics by external perturbations~\cite{4,5,6}. 2D materials thus are expected to make great impact on future science and technology. In graphene, most fantastic properties are associated with its nontrivial band topology, where the linear crossing between the valence and conduction bands can generate massless Dirac fermions at the Fermi level~\cite{7}. Benefiting from the fast developments on fabrication and exfoliation techniques~\cite{1,2,3}, a large number of 2D materials, ranging from the atomic silicene~\cite{8}, phosphorene~\cite{9,10}, borophene~\cite{11,12}, to binary (or polynary) transition-metal dichalcogenides~\cite{13} and MXenens~\cite{14}, have been successfully prepared in nowadays. Inspired by graphene, it has seen great interest in developing 2D materials with nontrivial electronic band structure.

In topological materials, the band crossing can potentially form zero-dimensional (0D) nodal point~\cite{15,16,yu1}, one-dimensional (1D) nodal loop~\cite{17,18,19,yu2} and 2D nodal surface~\cite{20,21,22,23}. Nodal loop materials have been hotly studied currently~\cite{24,25,26,27,28,29,30}, because of their intriguing properties such as the novel drumhead surface states and rich transport characteristics. Unlike the big number of nodal loop semimetals in 3D bulk materials, 2D nodal loop materials are much fewer~\cite{31,32,33,34,35,36,37}. Even more severely, these 2D nodal loop materials face one major challenge: almost all 2D nodal loops will be gapped in the presence of spin-orbit coupling (SOC). Recently, Wang \emph{et al.} predict the first 2D nodal loop semimetal robust against SOC in ferromagnetic (FM) material: MnN monolayer~\cite{38}. Magnetism can bring more possibilities in topological materials: the magnetic symmetry is highly dependent on the magnetization direction, thereby the topological states may change when the magnetization direction shifts. FM 2D LaCl monolayer is a such example, where the nodal loop can transform into Weyl semimetal or quantum anomalous Hall states under different magnetization directions~\cite{39}. For FM 2D nodal loop materials, it will be the most interesting if the nodal loop is fully spin-polarized, i.e., the nodal loop is formed by the bands from a single spin channel. Such fully spin-polarized nodal loop is highly desirable for the spintronics applications. Unfortunately, in most existing FM 2D materials such as LaCl monolayer~\cite{39}, CrAs$_2$ monolayer~\cite{40}, Na$_2$CrBi trilayer~\cite{41}, and GdAg$_2$ monolayer~\cite{42}, the nodal loops are only partially spin-polarized because the electron states from both spin channels coexist around the nodal loops. Considering these facts, exploring fully spin-polarized 2D nodal loop, especially those can robust against SOC, is highly desirable.

In nodal loop materials, the nodal loops, themselves, can even show different characteristics. For one example, they can take the form of Dirac and Weyl nodal loop based on the band degeneracy at the loop. The Weyl nodal loop requires the breaking of either inversion symmetry $\mathcal{P}$ or time-reversal symmetry $\mathcal{T}$~\cite{43,44,45}. For another example, nodal loops can show different slopes of band crossings, which classify them into type-I, type-II, and hybrid nodal loops correspondingly. The concept of hybrid nodal loop was been proposed quite recently~\cite{46,47}. Benefiting from the coexistence of type-I and type-II band crossings, hybrid nodal loop can exhibit quite unique magnetic responses such as zero-field magnetic breakdown and Klein tunneling~\cite{47}. Hybrid nodal loop has been predicted in a few nonmagnetic (NM) bulk materials~\cite{47,48,49,add1,50}, but has not been identified in 2D or magnetic system yet.

In this work, based on electronic structure calculations and symmetry analysis, we propose CrN monolayer as the first example of a 2D FM hybrid nodal loop semimetal. Interestingly, the hybrid nodal loop in CrN monolayer is fully spin-polarized and robust against SOC. The structure of CrN monolayer was initially prosed by Zhang \emph{et al.}, realized by the biomimetic particle swarm optimization~\cite{51}. They have demonstrated that CrN monolayer is both dynamically and thermally stable, and naturally show strong FM ordering with a high Curie temperature. Here, we reveal that CrN monolayer has a novel half-metallic electronic band structure, which features a fully spin-polarized Weyl nodal loop near the Fermi level. The nodal loop can exist even SOC is included, due to the protection of mirror symmetry. Remarkably, the nodal loop contains both type-I and type-II band crossings. Such hybrid nodal loop has not been identified either in 2D materials or in magnetic materials before. An effective Hamiltonian is constructed to describe the topological nature of the hybrid nodal loop. Most remarkably, we find the nodal loop can shift into either type-I or type-II Weyl nodes under in-plane magnetization, determined by the actual direction of magnetic field. Our findings promote a promising material platform to study the fundamental physics of hybrid Weyl nodal-loop, and type-I and type-II Weyl fermions in fully spin-polarized 2D system.

\section{COMPUTATIONAL DETAILS}

In this work, the first-principles calculations were performed by using the Vienna ab initio Simulation Package~\cite{52,53}. The exchange-correlation potential was adopted by the generalized gradient approximation (GGA) of Perdew-Burke-Ernzerhof (PBE) functional~\cite{54}. For the crystal structure of CrN monolayer, a vacuum with the thickness above 18 \AA $ $ was built to avoid potential interactions between layers. The long-range van der Waals interactions were taken into account by using the DFT-D2 method~\cite{55}. The cutoff energy was set as 600 eV. The Brillouin zone was sampled by a Monkhorst-Pack \emph{k}-mesh with size of 15$\times$15$\times$1. During our calculations, we applied the GGA+$U$ method to account for the Coulomb interaction of 3\emph{d} orbitals of Cr atom~\cite{56}. The effective \emph{U} value for Cr was chosen as 4 eV. To be noted, slightly shift the \emph{U} values will not change the main conclusion of our work.

\section{CRYSTAL STRUCTURE AND MAGNETIC CONFIGURATION}

\begin{figure}
\includegraphics[width=8.8cm]{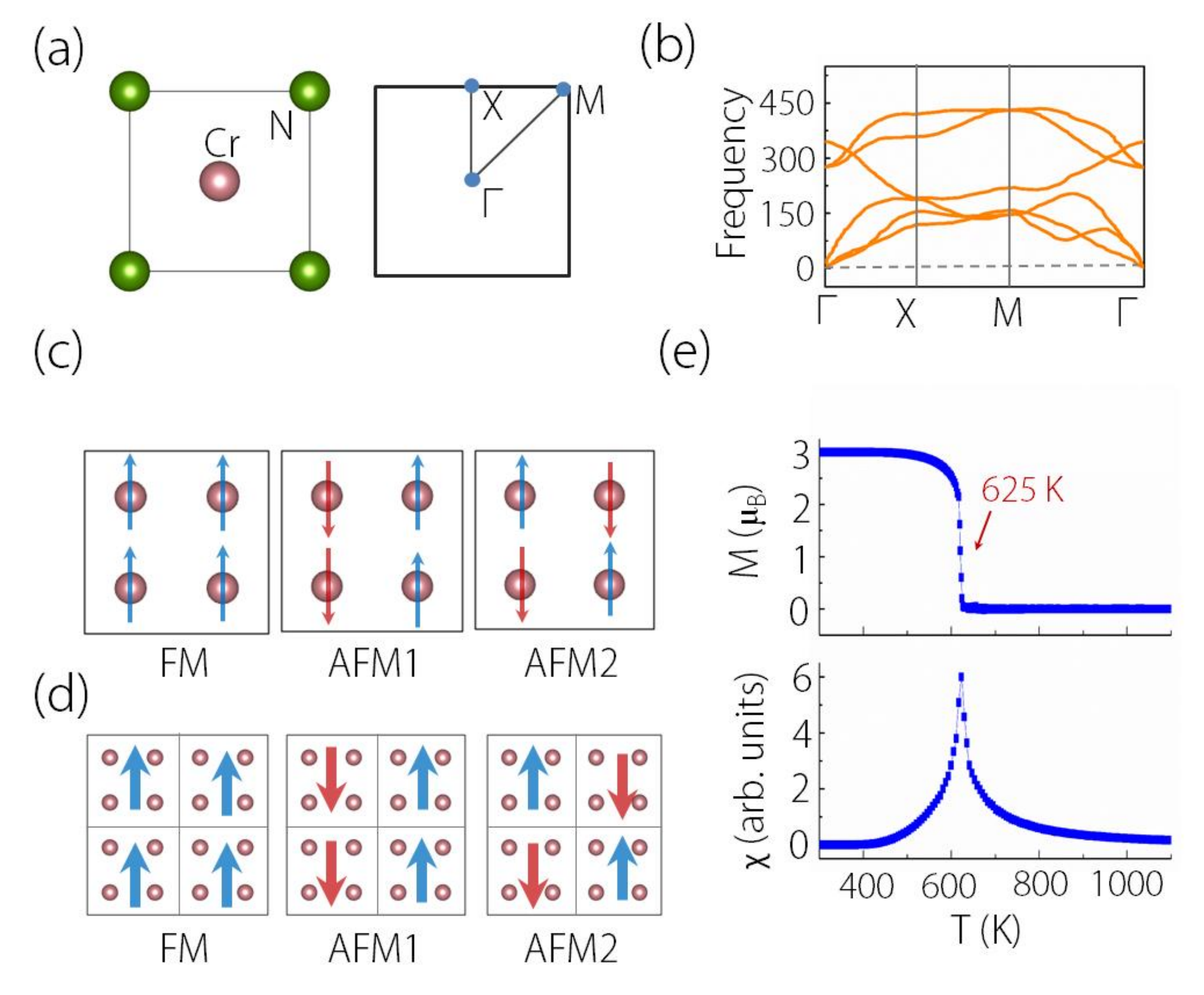}
\caption{(a) Primitive cell and 2D Brillouin zone of CrN monolayer. (b) Calculated phonon spectrum of CrN monolayer. (c) Three possible magnetic configurations (FM1, AFM1, and AFM2) in the 2$\times$2 supercell of CrN monolayer. Here, only the Cr atoms are shown in the supercell. The arrows denote the directions of spin. (d) is similar with (c) but for the 4$\times$4 supercell of CrN monolayer. (e) Simulated magnetic moment (\emph{M}) and magnetic susceptibility as a function of temperature.
\label{fig1}}
\end{figure}

The lattice structure of CrN monolayer is shown in Fig.~\ref{fig1}(a), featuring a 2D square lattice. It contains one Cr atom and one N atom in the primitive cell. The structure is completely flat with the Cr and N atoms situating in the same plane. The structure can be viewed as the monolayer form of the rocksalt CrN (001) surface. The structure takes the space group of \emph{P4/mmm} and the point group of \emph{D$_{4h}$}. To be noted, CrN monolayer possesses the mirror symmetry  $\mathcal{M}_{z}:(x,y,z)\rightarrow (x,y,-z)$, which is crucial for the discussions on its band topology. The optimized lattice constant is a = b = 2.825 \AA, in good agreements with the previous report (a = b = 2.82 \AA)~\cite{51}. In the optimized structure, the Cr-N bond length is 1.997 \AA. To recheck its dynamic stability, we have calculated the phonon spectrum of CrN monolayer. As shown in Fig.~\ref{fig1}(b), we find it shows no imaginary frequency mode throughout the Brillouin zone, suggesting CrN monolayer is dynamically stable.

Because of the partially occupied 3\emph{d} shell, the transition metal element Cr usually carry magnetic moments. We have considered different magnetic configurations including NM, FM and antiferromagnetic (AFM) ordering to identify the ground magnetic state in CrN monolayer. As shown in Fig.~\ref{fig1}(c), there exist two possible AFM configurations (AFM1 and AFM2) in the 2$\times$2 supercell. We compare the energy among these magnetic states, and find the FM state is the most energetically stable, while the NM and AFM states are respectively higher in about 1.52 eV and 0.64 eV per unit cell. We have also checked our results by applying the 4$\times$4 supercell [Fig.~\ref{fig1}(d)], where the FM state still possesses the lowest energy among all magnetic configurations. In the FM state, the total magnetic moment is 3.0 $\mu_B$, which is mostly contributed by the Cr atom.

To determine the easy axis in the FM state, we have calculated the magnetic anisotropy energy by scanning different magnetization directions in the presence of SOC. Our calculations show the out-of-plane [001] magnetization is most energetically stable, with about 0.1 meV lower than the in-plane configurations. These results have clarified that, CrN monolayer has a FM ground magnetic state with the magnetization along the [001] direction. Here, we also want to point out that, the energy difference between the out-of-plane and in-plane magnetizations quite close, thus the magnetic configurations can be easily tuned under external magnetic field.

The Curie temperature is a crucial parameter for FM materials, since it determines the stability of FM ordering. Here, we estimate the Curie temperature of CrN monolayer by performing the Monte Carlo simulation on the basis of Ising model~\cite{57}. The Hamiltonian for this 2D system can be expressed as:
\begin{equation}
{\mathcal{H}} = -\sum \limits_{ij}J_{ij}M_{i}M_{j},
\end{equation}
where $J_{ij}$ denotes the nearest-neighboring exchange parameter, and M represents the magnetic moment on the Cr site. As calculated by the energy difference between FM and AFM states, the exchange parameter \emph{J} for CrN monolayer is estimated to be 4.52 meV. To weaken the periodic constraints, we have performed a 100$\times$100 supercell in our Monte Carlo simulation. We show the curves of average magnetic moment and magnetic susceptibility versus temperature in Fig.~\ref{fig1}(e). We find the Curie temperature for CrN monolayer can be as high as 625 K. This value is much higher than typical 2D FM materials including Cr$_2$Ge$_2$Te$_6$ sheet ($\sim$ 66 K)~\cite{58}, Fe$_3$GeTe$_2$ layer ($<130$ K)~\cite{59}, MnN monolayer ($\sim$ 200 K)~\cite{38,60}, and metal-chlorides mnolayers (198-319 K)~\cite{61,62}. The high Curie temperature highly promise its application in spintronics.

\section{FULLY SPIN-POLARIZED HYBRID NODAL LOOP AND STRAIN-INDUCED TRANSFORMATIONS}

\begin{figure}
\includegraphics[width=8.8cm]{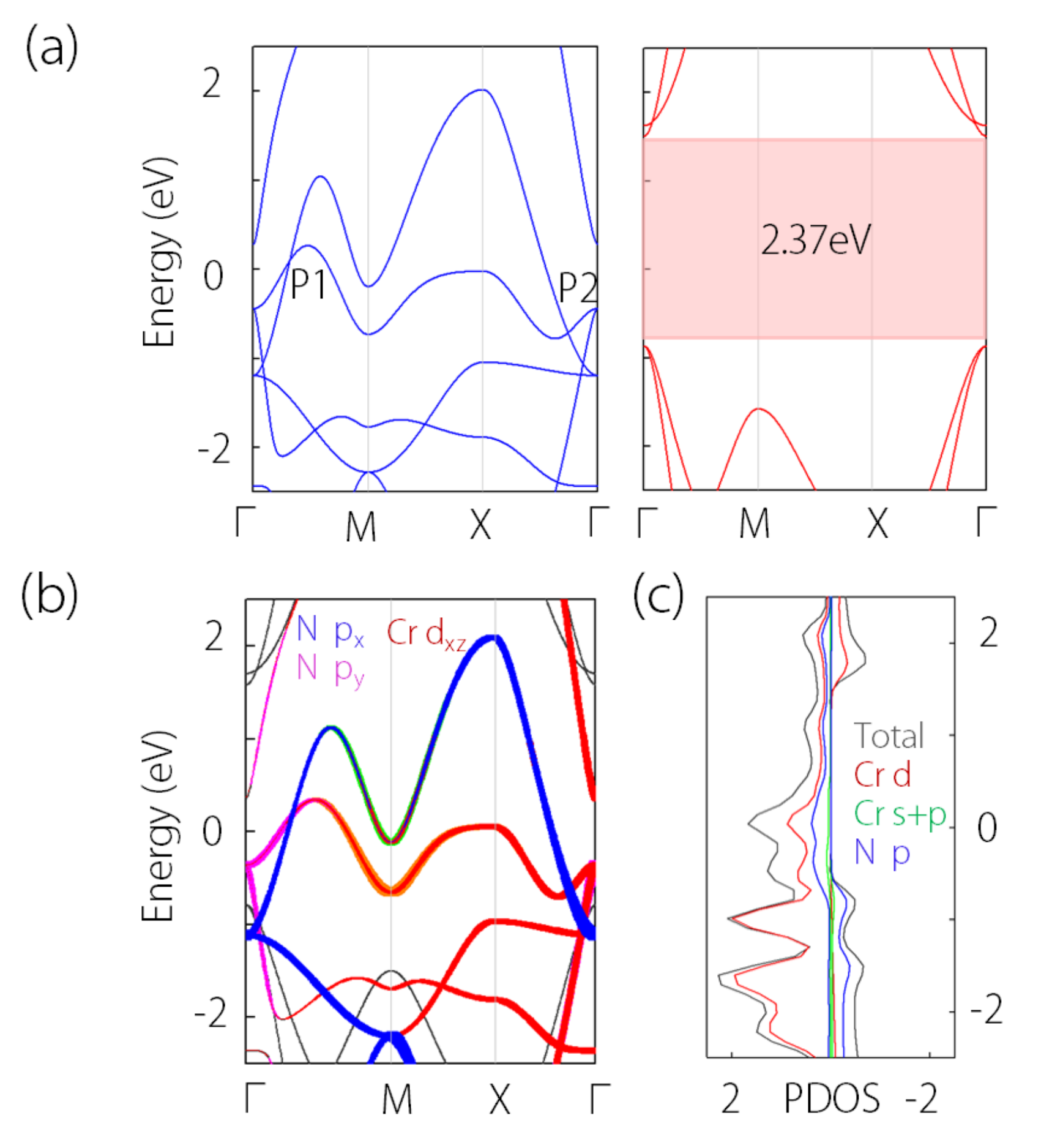}
\caption{(a) Electronic band structure of CrN monolayer in the spin-up (blue lines) and the spin-down (red lines) channels. In the spin-up channel, the band crossing points in the $\Gamma$-M and X-$\Gamma$ paths are denoted as P1 and P2. In the spin-down channel, the size of band gap is indicated. (b) The projection of band structure onto atomic orbitals including \emph{d} orbitals ($d_{xz}$) of Cr atom and \emph{p} orbitals ($p_{x}$, $p_{y}$) of N atom. (c) The total density of states (DOS) and orbital resolved partial DOS of CrN monolayer.
\label{fig2}}
\end{figure}

The most attractive feature of CrN monolayer lies in its electronic band structure. We first investigate the band structure without considering SOC. As shown in Fig.~\ref{fig2}(a), it clearly exhibits a half-metallic band structure. In the spin-up channel, it shows a metallic behavior with two bands crossing the Fermi level. However, the bands in the spin-down channel show a big gap of about 2.37 eV. Therefore, the conducting electrons in CrN monolayer are fully spin-polarized. In Fig.~\ref{fig2}(b), we show the orbital-projected band structure and density of states of CrN monolayer. We can find that, the states bands near the Fermi level are mostly contributed by the d orbitals ($d_{xz}$) of Cr atom and p orbitals ($p_{x}$, $p_{y}$) of N atom.

\begin{figure}
\includegraphics[width=8.8cm]{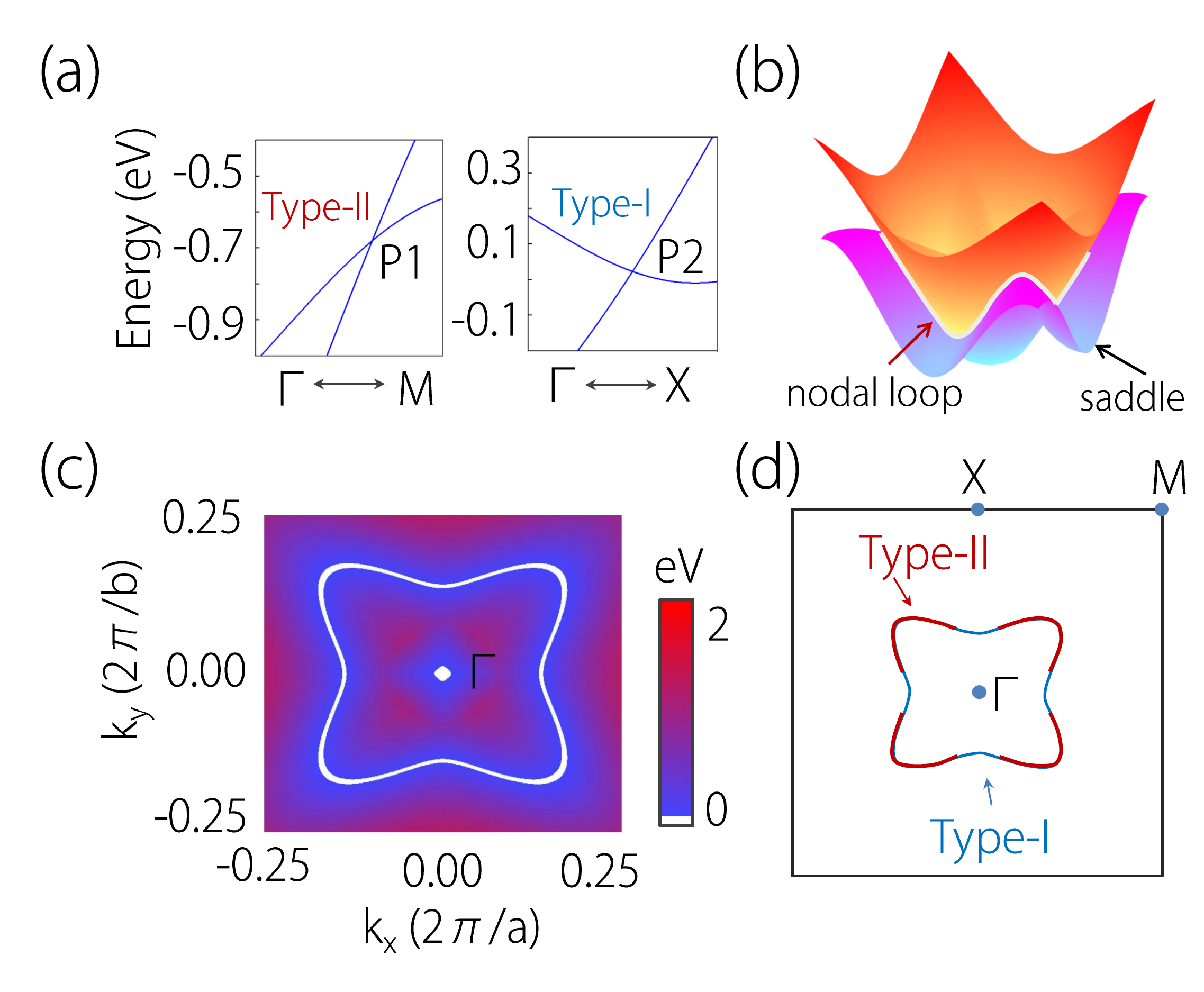}
\caption{(a) The enlarged view of band structures near P1 and P2 points. (b) Band dispersion around the $\Gamma$ point in the 2D Brillouin zone, showing the hybrid nodal loop. One of the crossing bands has a saddle-like shape. (c) The shape of hybrid nodal loop in the Brillouin zone. (d) shows the regions of type-I and type-II band crossings on the hybrid nodal loop.
\label{fig3}}
\end{figure}

Interestingly, we find the two bands near the Fermi level in the spin-up channel cross with each other, and produce two crossing points [P1 and P2, see Fig.~\ref{fig2}(a)]. Near P1 and P2, the two bands are inverted, as clearly shown by the orbital-projected band structure in Fig.~\ref{fig2}(b). Figure~\ref{fig3}(a) shows the enlarged view of band structures near P1 and P2. Even more interestingly, we find these crossing points possess different slopes of band dispersion, where P1 shows a type-I band crossing while P2 shows a type-II one. By further examination on the band structure, we find crossing points P1 and P2 are not isolate but reside on a nodal loop centering at the $\Gamma$ point. Figure~\ref{fig3}(b) shows the dispersion of the two crossing bands throughout the 2D Brillouin zone. We can clearly observe the presence of nodal loop. The profile of the nodal loop is shown in Fig.~\ref{fig3}(c). It is worth noticing that, the nodal loop in CrN monolayer is in fact a hybrid nodal loop, which hosts both type-I (such as P1) and type-II (such as P2) crossing points. The formation of hybrid nodal loop requires one of the crossing band is saddle-shaped~\cite{47}, which is indeed the case for CrN monolayer [see Fig.~\ref{fig3}(b)]. In Fig.~\ref{fig3}(d), we show the sections of type-I and type-II crossings on the nodal loop. To our best knowledge, this is the first proposal of a hybrid nodal loop in both 2D materials and magnetic materials.

To capture the nature of the nodal loop, we construct an effective \emph{k$\cdot$p} model. At the $\Gamma$ point, the two bands forming the hybrid nodal loop belong to the $\Gamma_{2}^{+}(A_{2g})$ and $\Gamma_{2}^{-}(A_{2u})$ representations of the \emph{D$_{4h}$} point group symmetry. Using them as the basis, we can establish an effective Hamiltonian (up to four order), expressed as
\begin{equation}\label{FNRm}
\mathcal{H}=\left[
              \begin{array}{cc}
                h_{11} & 0 \\
                0 & h_{22} \\
              \end{array}
            \right],
\end{equation}
with,
\begin{equation}
h_{ii}=M_{i}+a_{i}k^{2}+b_{i}(k_{x}^{4}+k_{y}^{4})+c_{i}k_{x}^{2}k_{y}^{2}
\end{equation}
Here, the momentum \emph{k} is measured from the $\Gamma$ point, i = 1, 2, and   $k=\sqrt{k_{x}^{2}+k_{y}^{2}}$. The coefficients $M_{i}$, $a_{i}$, $b_{i}$ and $c_{i}$ are real and material-dependent parameters, which can obtained by fitting the DFT band structure.

\begin{figure}
\includegraphics[width=8.8cm]{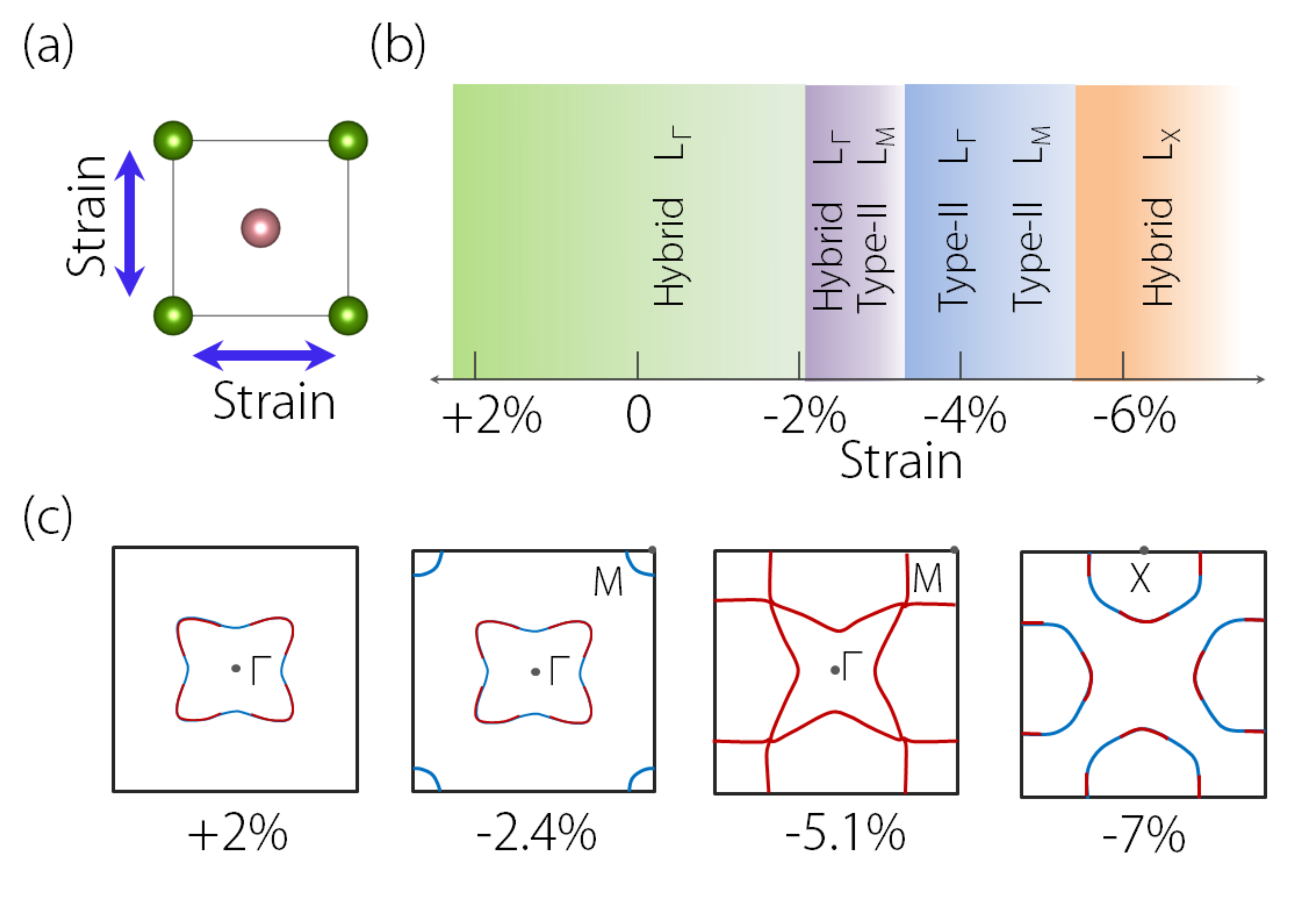}
\caption{(a) Illustration of applying biaxial in-plane strains on CrN monolayer. (b) Phase diagram of CrN monolayer under strain. (c) The profiles of nodal loop under specific strains (+2$\%$, -2.4$\%$, -5.1$\%$, and -7$\%$, where the signs ``+" and ``-" respectively denote the tensile and compressive strains). The sections of type-I and type-II band crossings are denoted as cyan and red colors, respectively.
\label{fig4}}
\end{figure}

	This effective Hamiltonian shows two key features. First, the two crossing bands for the nodal loop are decoupled, since the coupling parameters vanish in the Hamiltonian. This behavior arises from that, the system has a mirror symmetry with respect to the plane ($\mathcal{M}_z$), and the two bands with $\Gamma_{2}^{+}(A_{2g})$ and $\Gamma_{2}^{-}(A_{2u})$  representations possess opposite mirror eigenvalues, i.e. $n_{m}$ = -1 for the $\Gamma_{2}^{+}(A_{2g})$ band and $n_{m}$ = 1 for the other one. Second, if we keep the Hamiltonian up to the second order (by omitting the four-order terms), there exist a nodal loop when ($M_{1}$ - $M_{2}$)($a_{1}$ - $a_{2}$) $<$ 0 is satisfied. Specifically, the nodal loop is type-I for $a_{1}a_{2}$ $<$ 0 and type-II for $a_{1}a_{2}$ $>$ 0. For CrN monolayer, it satisfies $a_{1}a_{2}$ $>$ 0, and the nodal loop would be  type II  if the higher order terms are omitted. However, by including the higher order dispersion, we find while the existence of the nodal loop is not affected, as  it is protected by the mirror symmetry, the species of the nodal loop is transformed from type II to hybrid.

\begin{figure}
\includegraphics[width=8.8cm]{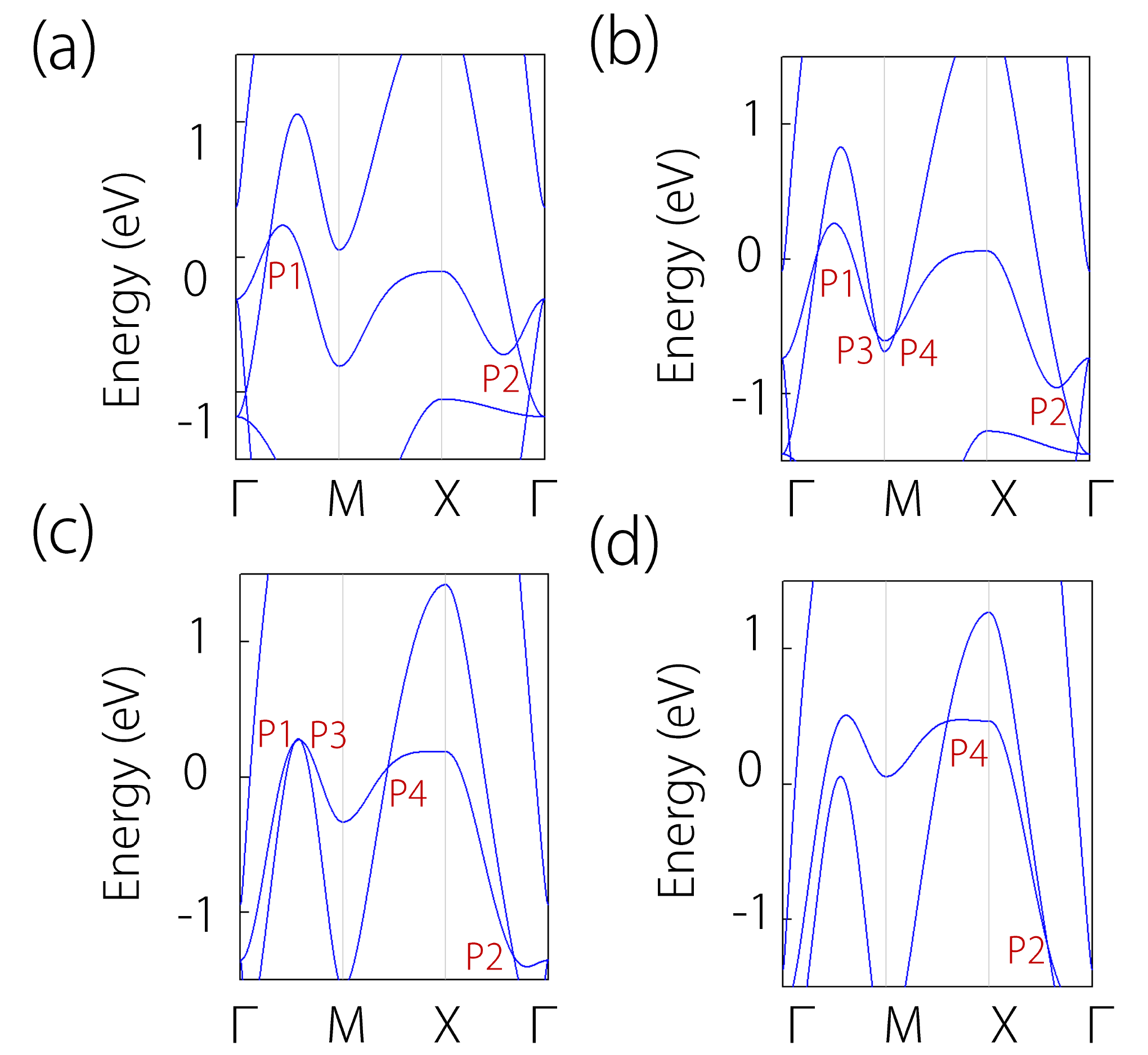}
\caption{Electronic band structure in the spin-up channel of CrN monolayer under (a) +2$\%$, (b) -2.4$\%$, (c) -5.1$\%$, and (d) -7$\%$ strains. In (a)-(d), the crossing points are indicated (P1, P2, P3, and P4).
\label{fig5}}
\end{figure}

\begin{figure}
\includegraphics[width=8.8cm]{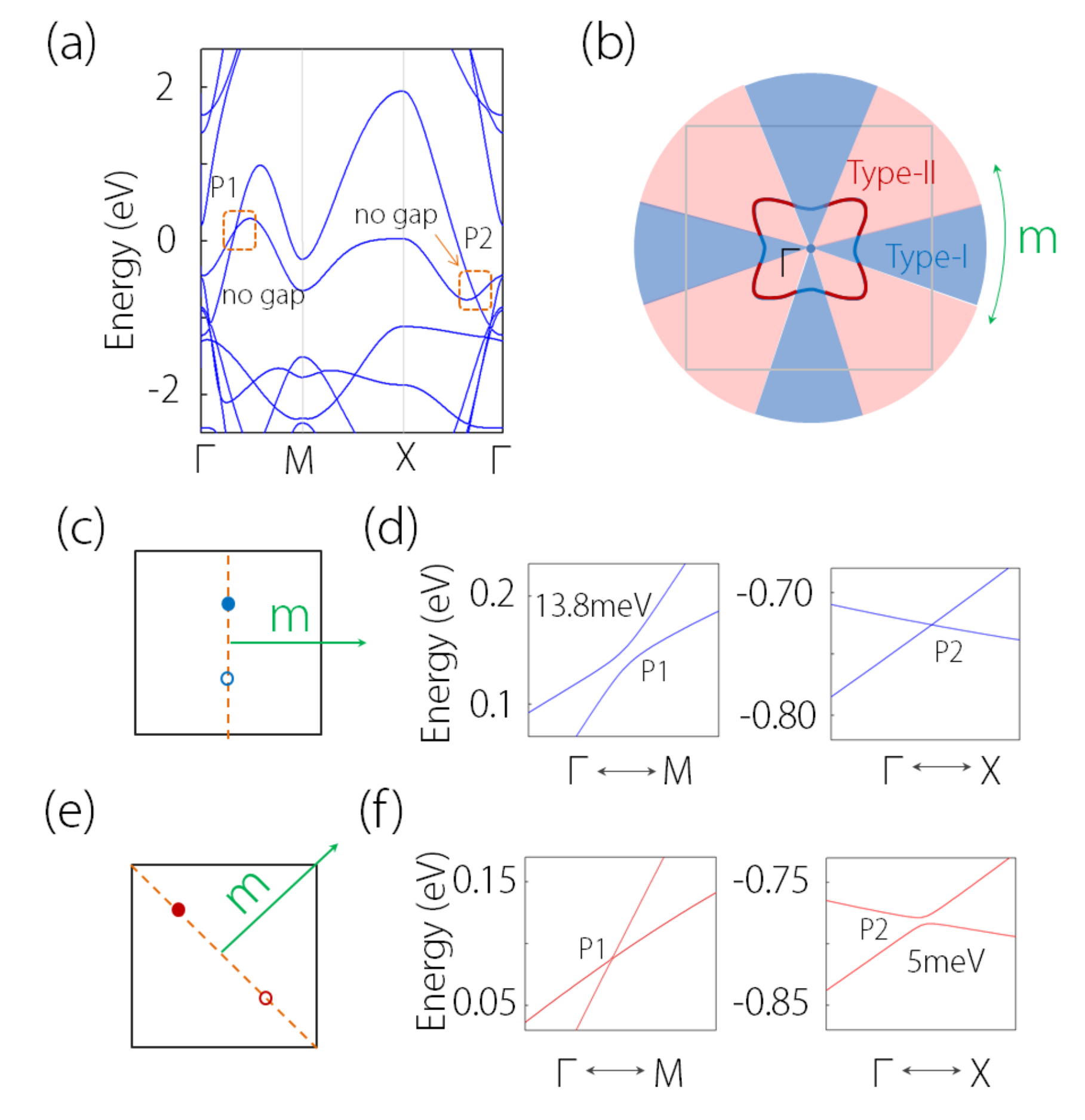}
\caption{ Electronic band structure of CrN monolayer under SOC with magnetization along the [001] direction. (b) The regions of type-I and type-II Weyl nodes on the hybrid nodal loop and the in-plane magnetizations. (c) The positions of the type-I Weyl nodes (the blue circles) under [100] magnetization. (d) The band structures near P1 and P2 under SOC with magnetization along the [100] direction. (e) and (f) are similar with (c) and (d) but for the [110] magnetization. In the [110] magnetization, a pair of type-II Weyl nodes (the blue circles) can be obtained in CrN monolayer, as shown in (e). In (c) and (e), the solid and hollow circles represent the ``positive" and ``negative" Weyl nodes.
\label{fig6}}
\end{figure}

In realistic materials, the electronic band structure are usually tunable under external perturbations such as strain. In CrN monolayer, we find the hybrid nodal loop shows interesting transformations under biaxial in-plane strains. To be noted, under biaxial in-plane strains [see Fig.~\ref{fig4}(a)], all the crystal symmetries including $\mathcal{M}_z$ will be retained. As shown in Fig.~\ref{fig4}(b) and (c), we can realize a rich phase diagram in CrN monolayer by such strain engineering. Under small tensile strains ($<$ 4$\%$), we find the band structure does not change much. Figure~\ref{fig5}(a) shows the band structure in the spin-up channel under a 2$\%$ tensile strain. We can observe that both P1 and P2 retain, where P1 shows a type-II band crossing while P2 shows a type-I. They produce a hybrid nodal loop centering the $\Gamma$ point (denoted as hybrid L$_{\Gamma}$) in CrN monolayer, as illustrate in Fig.~\ref{fig4}(b) and (c). However, the band structure can be significantly changed by applying compressive strains. For small compressive strains in the range of 2.3$\%$-3$\%$, CrN monolayer can show two nodal loops. One is a hybrid nodal loop centering the $\Gamma$ point (hybrid L$_{\Gamma}$), and the other is a type-II nodal loop centering the M point (denoted as type-II L$_{\rm M}$), as shown in Fig.~\ref{fig4}(b) and (c). Figure~\ref{fig5}(b) shows the band structure under a 2.4$\%$ tensile strain, where four crossing points (P1, P2, P3, and P4) can be found in the spin-up channel. Among them, P1 (which is type-II) and P2 (which is type-I) form the hybrid loop L$_{\Gamma}$, and the type-II crossing points P3 and P4 form the type-II loop L$_{\rm M}$. For larger compressive strains (3$\%$-5.2$\%$), we find the hybrid nodal loop L$_{\Gamma}$ will transform into a type-II one. Figure~\ref{fig5}(c) shows the band structure under a 5.1$\%$ tensile strain. We can observe that the four crossing points are all type-II. As result, CrN monolayer hosts two type-II nodal loops (L$_{\Gamma}$ and L$_{\rm M}$) in this occasion, as shown in Fig.~\ref{fig4}(b) and (c). For even larger compressive strains ($>$ 5.2$\%$), as shown in Fig.~\ref{fig4}(b) and (c), the two nodal loops touch with each other and transform into a new hybrid nodal loop centering the X point (denoted hybrid L$_X$). This transformation can be clearly shown in the band structure under a 7$\%$ tensile strain, as shown in Fig.~\ref{fig5}(d). In this case, the cross points P1 and P3 annihilate, and the remaining P2 (type-II) and P4 (type-I) form the hybrid L$_X$. These results suggest CrN monolayer is an excellent platform to investigate the novel nodal loop transformations in 2D system with time-reversal symmetry breaking.

\section{TUNABLE WEYL FERMIONS FROM HYRID NODAL LOOP}

As discussed above, the hybrid nodal loop in CrN monolayer is protected by the mirror symmetry $\mathcal{M}_z$. For most of the previous studies on nodal loops semimetals protected by mirror symmetry, the nodal loop vanishes when SOC is included~\cite{31,32,33,34,35,36,37}. In contrast, we find the nodal loop in CrN monolayer can be robust against SOC. Here, the nodal loop is formed by two bands in same spin channel with opposite mirror eigenvalues $n_{m}$ = $\pm$1. Under SOC, the mirror symmetry is retained, because of the out-of-plane [001] spontaneous magnetization. Thereby, the mirror eigenvalues of the two bands keep opposite, although the eigenvalues have changed into $n_{m}$ = $\pm$i. This shows that, the two crossing bands are still decoupled and the hybrid nodal loop can survive under SOC. Such a scenario has been verified by our DFT calculations. Fig.~\ref{fig6}(a) shows the band structure of CrN monolayer under SOC. We can observe that the bands for two spin channels couple together, but the crossing points P1 and P2 on the nodal loop are not gapped. We have also checked other parts of the nodal loop, and found no SOC gap.

It is well known that, the magnetic symmetry highly depends on the magnetization direction, thus the topological state can be changed if the magnetization direction shifts. The CrN monolayer has a ``soft" magnetism because the energies among all potential magnetic configurations are very close, which suggests its magnetic configurations can be easily tuned under external magnetic field. This  fact makes it promising to realize distinct signatures under different magnetic fields. We show this point by applying in-plane magnetizations in CrN monolayer. Under in-plane magnetizations, the mirror symmetry $\mathcal{M}_z$ is broken, so that the hybrid nodal loop should vanish correspondingly. We show the band structures under the [100] and [110] magnetization directions in Fig.~\ref{fig6}(d) and (f), respectively.

For the [100] magnetization, we find the crossing point P1 is gapped with 13.8 meV but P2 is retained [see Fig.~\ref{fig6}(d)]. We have also checked other parts of nodal loop, and obtained no other crossing points. Therefore, CrN monolayer has transformed into a type-I Weyl semimetal with a single pair of time-reversal breaking Weyl nodes in the $\Gamma$-X path [see Fig.~\ref{fig6}(c)]. Very interestingly, we find CrN monolayer can further transform into a type-II Weyl semimetal under the [110] magnetization direction, as shown in Fig.~\ref{fig6}(e) and (f). In Fig.~\ref{fig6}(b), we again show the regions of type-I or type-II parts on the hybrid nodal loop. In fact, the presence of Weyl nodes for the in-plane magnetization can be illustrated in the symmetry view. Here, we use the [110] magnetization as an example (the symmetry analyses for [100] magnetization is fundamentally the same). With the [110] magnetization, the system preserves the magnetic double point group \emph{C$_{2h}$}. The little group along the $\Gamma$-M direction is \emph{C$_{2}$}, which can allow band crossing along this direction, thereby resulting the type-II Weyl nodes in the $\Gamma$-M path [see Fig.~\ref{fig6}(e)].

\section{DISCUSSION AND CONCLUSION}

\begin{figure}
\includegraphics[width=8.8cm]{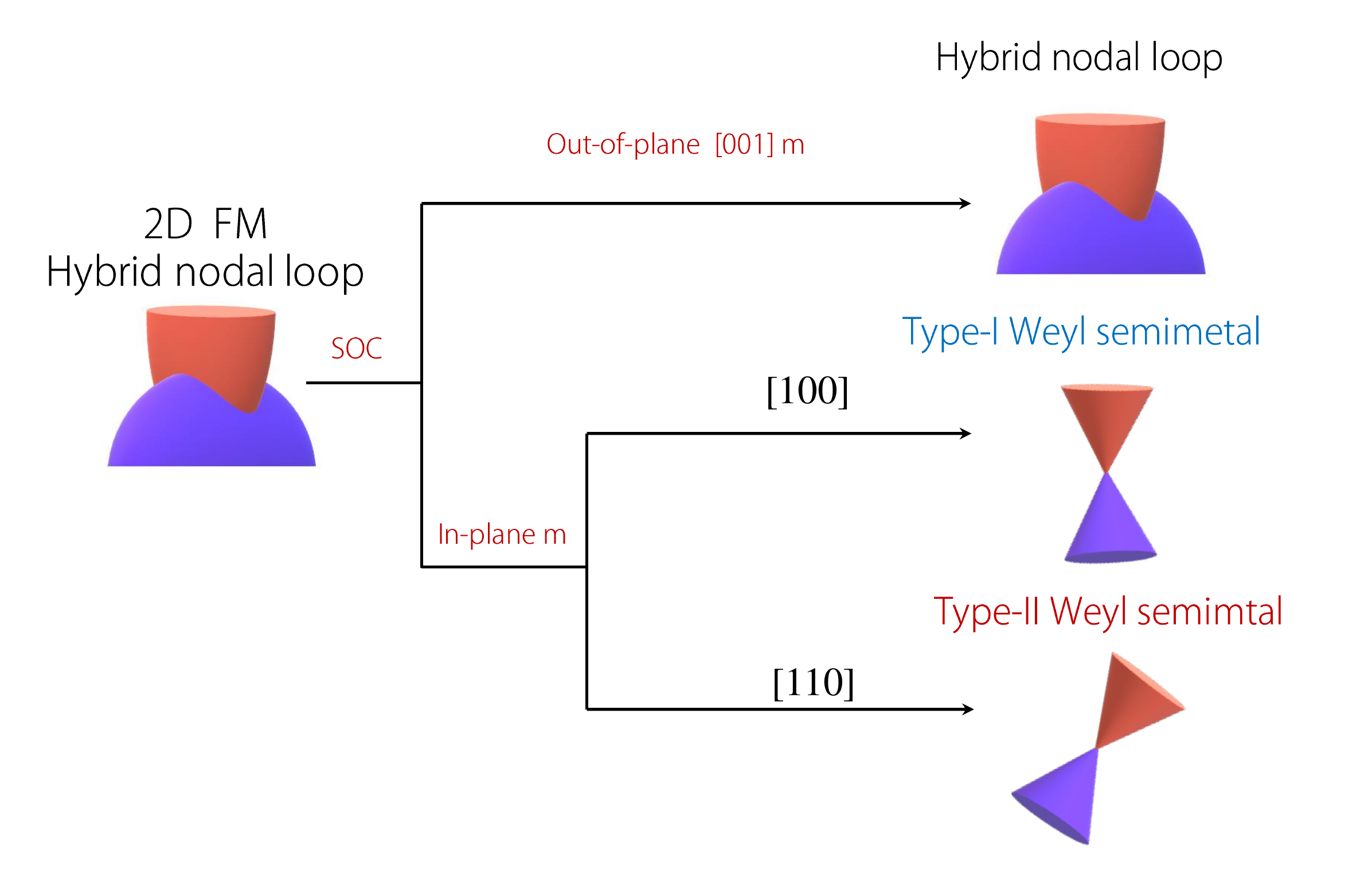}
\caption{Schematic diagram of different topological phases realized in CrN monolayer. These topological phases depend on the direction of magnetization, the mirror symmetry and the SOC.
\label{fig7}}
\end{figure}

Before closing, we have following remarks. First, our most important finding in this work is that, we proposes the first example of hybrid nodal loop both in 2D materials and in magnetic system. To date, hybrid nodal loop is only proposed in a few 3D materials including tetragonal Ca$_2$As~\cite{47}, Li$_2$BaSi~\cite{48}, CsCl-type YCd~\cite{49}, CrP$_2$O$_7$ compound ~\cite{add1} and binary Be$_2$Si~\cite{50}, and has been identified in 2D materials yet. Moreover, these examples are almost non-magnetic materials, i.e., their nodal loops are contributed by electron states from both spin channels. However, the hybrid nodal loop in CrN monolayer is fully spin-polarized, which is formed by bands from a single spin channels. Such hybrid nodal loop is not only desirable for potential spintronics application, but also can realize rich Weyl states under external magnetic fields, which is described as a Weyl nodal loop under out-of-plane [001] magnetization and either type-I or type-II Weyl nodes under in-plane magnetizations, as illustrated in Fig.~\ref{fig7}. Considering this, the fully spin-polarized hybrid nodal loop in CrN monolayer plays a good platform to investigate variable FM Weyl fermions.

Second, the nodal loop in CrN monolayer is strictly protected by symmetry, and is robust against SOC. Almost all the nodal loops in 2D materials reported so far will be gapped when SOC is included. Especially when SOC gaps are not neglectable in heavy-elements-containing materials, they cannot be described as 2D nodal loop semimetals. Hexagonal MnN monolayer is an excellent example for 2D nodal loop semimetal with hosting two nodal loops robust against SOC~\cite{38}. Both nodal loops in this example show type-I band dispersion, while our work tells a different story because it shows a hybrid loop with the coexistence of both type-I and type-II band crossings.

Finally, the nodal loop in CrN monolayer also have several additional features: (i) it is a time-reversal breaking Weyl loop, because the material is FM but preserves the inversion symmetry; (ii) it can transform into other nodal loop configurations under external perturbations, such as the type-II case under expressive strain [see Fig.~\ref{fig4}], which makes band topology of CrN monolayer even meaningful; (iii) the nodal loop band structure is fairly clean without coexisting of other extraneous bands, thus the hybrid nodal loop here is very promising to be detected in experiments.

In conclusion, we have revealed 2D CrN monolayer as the first FM hybrid nodal-loop semimetal. Our results show CrN monolayer has a good stability, and exhibits a FM ground state with the Curie temperature as high as 675 K. The material shows a half-metallic band structure, and features a fully spin-polarized nodal loop in the spin-up channel. The nodal loop is robust against SOC, under the protection of mirror symmetry. Very interestingly, we find the nodal loop shows both type-I and type-II band crossings, which is known as a hybrid nodal loop. Such hybrid nodal loop has not been identified in 2D materials or in magnetic materials before. We have built an effective model, which can well characterize the emergence of hybrid nodal loop in the material. We also show that, both the band slope and the geometric configuration of the nodal loop can be shifted under lattice strain. Most remarkably, we find the hybrid nodal loop can further transform into discrete type-I and type-II Weyl nodes under in-plane magnetizations, which makes it feasible to design type-I and type-II Weyl fermions in a single material. Our work provides an excellent platform to study the intriguing properties of hybrid nodal loop and Weyl fermions in 2D materials with time-reversal breaking.

\begin{acknowledgments}
This work is supported by National Natural Science Foundation of China (Grants No. 11904074), Nature Science Foundation of Hebei Province (Nos. E2019202222 and E2019202107) and Beijing Institute of Technology Research Fund Program for Young Scholars. One of the authors (X.M. Zhang) acknowledges the financial support from Young Elite Scientists Sponsorship Program by Tianjin.
\end{acknowledgments}

\end{document}